\documentclass[11pt,a4paper]{article}

\usepackage{amsthm,amsmath}
\usepackage{graphicx} 
\usepackage{setspace} 
\usepackage[top=70pt, bottom=80pt, left=80pt, right=70pt]{geometry}

\usepackage{multirow}
\usepackage{amssymb}
\usepackage{amsmath}
\usepackage{float}
\usepackage{etoolbox}
\usepackage{lineno,hyperref}
\modulolinenumbers[5]
\usepackage[algo2e]{algorithm2e} 
\usepackage[noend]{algpseudocode}
\usepackage{algorithm}
\usepackage{color}

\AtEndEnvironment{pf}{\hfill\null $\square$}

\begin{document}

\begin{center}

{\large \bf Approximation Algorithms for Multi-Multiway Cut and Multicut Problems on Directed Graphs}

Ramin Yarinezhad$ ^1 $, Seyed Naser Hashemi$ ^1 $ \footnote[2]{ *Corresponding Author}$ ^* $

$ ^1 $Department of Mathematics and Computer Science, Amirkabir University of Technology, Tehran, Iran

\end{center}

\doublespacing

\noindent{\bf Abstract}

In this paper, we present two approximation algorithms for the directed multi-multiway cut and directed multicut problems. The so called region growing paradigm \cite{1} is modified and used for these two cut problems on directed graphs. 
By using this paradigm, we give for each problem an approximation algorithm such that both algorithms have the approximate factor $O(k)$ the same as the previous works done on these problems. However, the previous works need to solve $k$ linear programming, whereas our algorithms require only one linear programming. Therefore, our algorithms improve the running time of the previous algorithms. 

\noindent{\bf Keywords}

\noindent
Approximation algorithm, Computational complexity, NP-hard problems, Directed multi-multiway cut, Directed multicut cut

\noindent
\textbf{2010 MSC:} Primary: XXXXX; Secondary: XXXXX, XXXXX
\vspace{13cm}

\pagebreak

\section{Introduction} \label{section.intro}
In the following, we first review some of the important cut problems which serve as a background for the problems considered in this paper.
The undirected multiway cut problem is defined on an undirected graph $G=(V,E)$ with a given set $S=\{s_1,...s_k\} \subseteq V$ of vertices called terminals and a weight function $ c_e,\ e \in E $. Here, the goal is to find the minimum weight subset of edges so that by deleting them, all terminals in $S$ are disconnected. In other words, there is not any path between any two considered terminals. It is proved that this problem, for $ k \geq 3$, is NP-hard and MAX SNP-hard, for which a $2-2/k$ factor approximation algorithm is given \cite{5}. In \cite{6}, using a geometric relaxation, an algorithm with an approximate factor of $1.5-1/k$ is introduced and it is improved to $1.3438- \epsilon _k $ in \cite{7}.

For directed graphs, the version of the directed multiway cut problem is defined. Likewise, given a set of terminals $S=\{s_1,...s_2\} \subseteq V$ and a weight function $ c_e,\ e \in E $, we look for a minimum weight subset of edges whose deletions disconnect all directed paths between each pair of terminals. Vazirani and Yannakakis \cite{3} showed that a directed multiway cut problem is also NP-hard and MAX SNP-hard. They introduced an algorithm with a $2 \log k $ approximate factor. The best known approximation algorithm, presented by Noar and Zosin \cite{4}, used a novel relaxation multiway flow to have an approximation algorithm within a factor of 2.

The problem of undirected multicut is another well-known problem defined on undirected graphs with a non-negative cost $ c_e,\ e \in E $, and a set of ordered pairs of vertices, namely; $ (s_1,t_1),...,(s_k,t_k)$, which are called source-terminal vertices. In this case, the seek is to achieve a minimum cost subset of edges so that removing them all sources become inaccessible from their corresponding terminals. For $ k \geq 3 $ , it is shown that the problem is NP-hard and MAX SNP-hard \cite{5}. Garg, Vazirani, and Yannakakis \cite{1} give, by the region growing technique, an approximation algorithm with the $O(\log k)$ approximate factor. In \cite{18} for this problem with more constraints, an approximation algorithm has been proposed with approximation factor $O(rlog^{3/2}k)$, where $r$ is a part of the input instance.

The directed multicut problem is defined as follows: given a directed graph $ G=(V,E) , |V|=n$ with a non-negative function $c_e>0, e \in E$, and a set of ordered pairs of vertices $(s_1,t_1),...,(s_k,t_k)$, we find a subset $F \subseteq E $ with minimum cost function so that their removal from the graph makes each pair disconnected. That is, for any $i, 1  \leq i  \leq k, $ there is not any directed path from $s_i$ to $t_i$ in the graph $G(V, E-F)$. 

Furthermore, if the desire is also to disconnect the paths from $t_i$ to $s_i$, for any $i, 1 \leq i  \leq k$, we have an alternative version of the multicut problem called directed symmetric multicut problem. 

As shown in \cite{3}, for $ k \geq 2 $, the directed multicut problem is NP-hard and MAX SNP-hard. In some papers it was shown that another version of this problem is NP-hard \cite{13}. In literature, most of the works on directed multicut have been focused on the directed symmetric multicut problem \cite{9,10,11,12}. Even, Noar, Schieber and Sudan \cite{11} presented an approximation algorithm with a factor $ O((log)loglogk)$. In general, for a non-symmetric version, using the technique of region growing, an algorithm with the approximate factor $O(\sqrt{nlogk})$ is given \cite{14}. For the general case, Gupta \cite{15} introduced a simpler algorithm and improved the approximate factor to $O(\sqrt{n})$. Both problems above, studied by \cite{14,15}, use a linear programming relaxation to approximate the solution. In the work of Saks, Samorodnitsky, Zosin \cite{17}, it is shown that the integrality gap for the linear programming relaxation is $O(k)$.

A more general problem on undirected graphs is the multi-multiway cut problem in which the weight function $ w:E \to \Re^+$ and k sets $S_1,S_2,...,S_k$ are given. Here, our aim is to obtain a minimum weight subset of edges whose removal from the graph will disconnect all connections between the vertices in each set $S_i$, for $1 \leq i \leq k$. 
For $K=1$, this problem is an undirected multiway cut problem, and if $|S_i|=2, (1\leq i \leq k)$, an undirected multicut problem is obtained. Avidor and Langberg \cite{2} showed that the undirected multi-multiway cut problem is NP-hard and MAX SNP-hard, and by using the region growing technique they could present an approximation algorithm within the factor of $O(logk)$. When the input graph is a tree, in \cite{16} has been shown that this problem is solvable in polynomial time, if the number of terminal sets is fixed and in \cite{8} has been presented an approximation algorithm with a factor $ O(\sqrt k)$.

A directed version of the above problem is also defined namely as a directed multi-multiway cut problem. Similarly, for this problem a weight function $ w:E \to \Re^+$ on edges and $k$ sets  $S_1,S_2,...,S_k$ are  given. We seek to find a minimum weight subset of edges whose removal from the graph will disconnect all paths between the vertices in each set $S_i$, for $1 \leq i \leq k$. This problem generalizes the problems of directed multiway cut and directed symmetric multicut (when $k=1$ and $|S_i|=2$, respectively).

Since every instance of the directed multiway cut is defined as an instance of the directed multi-multiway cut problem when $k=1$, so the hardness proof for the multiway cut problem implies that the directed multi-multiway cut problem is also NP-hard and MAX SNP-hard.

Note that the problem of directed multi-multiway cut cannot be viewed as a generalization of the undirected multi-multiway cut problem only by replacing each undirected edge by two unparalleled directed edges. For example, consider a tree with a root $r$, containing three leaves $a,b,c$, and assuming the weight of each edge is equal to one. In this case, we get the optimal value, $OPT = 2$, whereas substituting each edge by two directed edges gives $OPT=3$, and this proves that two problems above are not equivalent.

As described above, it is clear that the problems of directed multicut and directed multi-multiway cut can be approximated by a factor $O(k)$. But for each of these problems, we require $k$ linear programming to be solved in order to obtain the desired approximation solution. In this paper, we show that we can achieve the same result, i.e. an approximation with the factor $O(k)$, by solving only one linear programming. To achieve this goal, the so called paradigm of region growing, introduced in \cite{1} for undirected cut problems, is modified so that it can be useful to produce an approximate solution of the multicut and multi-multiway cut problems on directed graphs.

\subsection{Organization}
The rest of this paper is organized as follows: In section \ref{section.lpr}
we present a linear programming relaxation for the Directed Multi-Multiway Cut problem which is used in \cite{2} and \cite{3}. Section \ref{section.def} contains necessary definitions and lemmas for the algorithm Directed Multi-Multiway cut which proposed in section \ref{section.dmmc}. Directed Multicut Algorithm presented in section \ref{section.dm}
and Conclusion is brought in section \ref{section.con}.

\section{Linear Programming Relaxation for the Directed Multi-Multiway Cut} \label{section.lpr}
We define a decision variable $ x(e) $ for each edge $e$ which is as follows: if $e$ belongs to directed multi-multiway cut, $x(e)=1$, otherwise $x(e)=0$. The purpose is to find a directed multi-multiway cut with the minimum cost which disconnects every directed path between two vertices in a group. We call that the set of all directed paths between any two vertices belongs to a group, with $P$. An integer program for the problem is given by:

\begin{equation*} 
	\begin{array}{ll@{}ll}
		minimize  & \displaystyle\sum\limits_{e \in E} & w(e)x(e) &\\
		$\textit{subject} $ $\textit{to}$ & \displaystyle\sum\limits_{e \in p} &x(e) \geq 1,  & \forall p \in P \\
		&                                                &x(e) \in \{0,1\}, & \forall e \in E
	\end{array}
\end{equation*}

By relaxing this IP we obtain the following linear programming relaxation:

\begin{equation*} 
	\begin{array}{ll@{}ll}
		minimize  & \displaystyle\sum\limits_{e \in E} & w(e)x(e) &\\
		$\textit{subject} $ $\textit{to}$ & \displaystyle\sum\limits_{e \in p} &x(e) \geq 1,  & \forall p \in P \\
		&                                                &x(e) \geq 0, & \forall e \in E
	\end{array}
\end{equation*}

In this LP, there is a constraint for each path. On the other hand, we may have an exponential number of paths with respect to the input size and as a result, exponential number of constraints. Nevertheless, we can solve this LP in polynomial time, using the ellipsoid algorithm. For this LP, the separation oracle operates as follows: we get a solution $x$ and assume that the length of each edge $e$ is equal to $x(e)$. Then, we find the shortest directed path between two vertices which are needed to be disconnected from each other. For example $(u,v)$, if the shortest path between $u$ and $v$ (either $v \rightarrow  u$ or $u \rightarrow v$) is more than 1, then this constraint $ \sum_{e \in p} x(e) \geq 1 $ is true for all paths between these two vertices. Therefore, this LP can be solved in polynomial time. 

To express the approximation algorithm for directed multi-multiway cut, we need several definitions and lemmas which are presented in the next section.

To round the solution of the mentioned LP and obtain a directed multi-multiway cut, we use the region growing technique \cite{1,2}. Note that definitions in \cite{1,2} are related to undirected graphs while definitions presented here, are related to directed graphs. 

We define a distance on edges and assume $x$ is an optimal solution for the LP. Let $x(e)$ be the length of edge $e$. The distance between two vertices $u$ and $v$ (either $v \rightarrow  u$ or $u \rightarrow v$), which is defined based on $x(e)$, is the length of the shortest path between them. We represent this shortest path with $dist(u,v)$. If there is no directed path between two vertices $u$ and $v$, the value of $dist(u,v)$ is equal to the length of the shortest path in the graph, regardless of the direction of edges. We define:

\begin{align*} 
	B_x(s_{ij},r)=\{v \in V : dist(s_{ij},v) \leq r\},
\end{align*}

where $1 \leq i \leq k$, $1\leq j \leq |S_i|$ and $r\in \mathbb{R}$. $B_x(s_{ij},r)$ is an area like a ball with center $s_{ij}$ and radius $r$.	Assume that $ \delta(s)$ is the set of all edges which only one of their endpoints is in the set $s$. For a given radius $r$, let $wt(\delta(B_x(s_{ij},r)))$ be the sum of weights of all edges which one of their endpoints is in $B_x(s_{ij},r)$. $wt(\delta(B_x(s_{ij},r)))$ is defined as follows:

\begin{align*}
	wt(\delta(B_x(s_{ij},r)))=\sum_{e \in \delta(B_x(s_{ij},r))} w(e),
\end{align*}

where $w(e)$ is the weight of edge $e$. The same as \cite{2}, let $c_i(r)$ be the sum of weights of directed edges whose one head only is inside these balls, where $1 \leq i \leq k$. $c_i(r)$ is defined as follows:

\begin{align*}
	c_i(r)=\sum_{j=1}^{|S_i|} wt(\delta(B_x(s_{ij},r))).
\end{align*}

Assume that each edge $e$ in the graph as being a pipe with cross-sectional area $w(e)$ and length $x(e)$. Then, the product $w(e)x(e)$ is equal to the volume of edge $e$. 
Thus, the solution of LP is the minimum volume of edges such that $dist(u,v) \geq 1$, where $u$ and $v$ are in the same group, and there is a path between them (either  $v \rightarrow  u$ or $u \rightarrow v$). 
Let $x$ be an optimal solution for the LP, and $V^*=\sum_{e \in E} w(e)x(e)$ be the volume of all edges. We know that $V^* \leq OPT$ such that $OPT$ is the optimal value for the IP. $v_i(r)$ is defined as follows:

\begin{align*}
	v_i(r) = \beta V^* + 
	& \sum_{j=1}^{|S_i|} (\sum_{
		\substack{
			e=(u,v)\in E \\ u,v \in B_x(s_{ij},r)
		}
	} w(e)x(e)+ \sum_{
		\substack{
			e=(u,v) \in E \\  u \in B_x(s_{ij},r) \\ v \notin B_x(s_{ij},r)
		}
	} w(e)(r-dist(s_{ij},u))),
\end{align*}

where $\beta > 0$ and is independent from $r$. We notice that an edge may appear in $c_i(r)$ more than once. That means we may have 
$ \delta (B_x(s_{ij},r)) \cap \delta (B_x(s_{ij^{'}},r)) \neq \emptyset $, for $ 1 \leq j \neq j^{'} \leq |S_i| $. Thus, $c_i(r)$ is an upper bound on the cut. According to these definitions, we can express the Lemma 1, which is used in the proof of Lemma 2.

\textbf{Lemma 1:} The function $v_i(r)$ is differentiable in $(0,\infty)$ except some finite number of points. The derivative of this function is $c_i(r)$.

\textbf{Proof:} The function $v_i(r)$ is not differentiable in points which the value of function $B_x(s_{ij},r)$ changes. 
The function $B_x(s_{ij},r)$, changes for the values of $r$ in which there is a vertex $v$ such that $dist(s_{ij},r)=r$. Thus the number of points in which the function $v_i(r)$ is not differentiable, is finite. Beside this, according to the definition done for the function $v_i(r)$, the derivative of this function is $c_i(r)$.
\begin{flushright}
	$ \blacksquare $
\end{flushright}

Lemma 2 says in directed graphs, we can always find a radius $ r < \frac{1}{2} $, such that the cost $v_i(r)$ is an upper bound for $c_i(r)$.

\textbf{Lemma 2:} Let $x$ be a feasible solution for the LP, for every $s_{ij}$ there is a $ r < \frac{1}{2}$ and at least an $\alpha$ $(\alpha > 0)$ such that the following inequality is true:

\begin{center}
	$c_i(r) \leq \alpha v_i(r)$.
\end{center}

The proof of this lemma is given in the Section 4.1. We first present the algorithm using this lemma.

\section{Approximation Algorithm for Directed Multi-Multiway Cut} \label{section.dmmc}

Our polynomial time approximation algorithm for directed multi-multiway cut is described in Algorithm 1. The algorithm solves the LP first and finds the optimal solution $x$. Then, the algorithm enters to a repetition loop and till there exists a path between two vertices in a group, the algorithm works as follows: assume that the set $S_i$ is chosen in this iteration. 
In the beginning, it finds an $r$ which satisfies the inequality of Lemma 2, and then it finds the set of balls with the center of vertices inside the $S_i$ with the radius of $r$. Then it puts the edges, which have been cut by these balls, in the answer set $F$.

Now the algorithm checks whether there are two vertices from another group which are connected by a path and are in the same ball. If there are such vertices, only the vertex in the center of each ball and its incident edges will be deleted from the graph. Otherwise, all of the vertices in balls and incident edges with them will be deleted from the graph.

\begin{algorithm} [H] 
	\SetAlgoLined
	\KwResult{A Directed Multi-Multiway Cut }
	$F \leftarrow \emptyset $ \\
	Solve the LP and get the optimal solution $x$ \\
	\While{there is a path between $s_{ij} \in S_i$ and $s_{ij^{'}} \in S_i $, where $1 \leq i \leq k$ and $1 \leq j\neq j^{'} \leq |S_i|$ }{
		Find $r_i$ such that $c_i(r_i) \leq \alpha v_i(r_i)$\\
		Add $\bigcup_{j=1}^{|S_i|} \delta (B_x(s_{ij},r_i))$ to F\\	
		\eIf{$\bigcup_{j=1}^{|S_i|} B_x(s_{ij},r_i)$ contains two vertices $u,v$ such that $u,v \in B_x(s_{ij},r_i)$, where $1 \leq j \leq |S_i|$ AND $u,v \in S_m$, where $1 \leq m \leq k$ and $m \neq i$ AND there is a path between $u$ and $v$ }{
			Remove $s_{ij}$, where $1 \leq j \leq |S_i|$, and incident edges with it form the graph\\
		}{
			Remove $\bigcup_{j=1}^{|S_i|} B_x(s_{ij},r_i)$ and incident edges with it form the graph\\
		}
		$\forall l \in \{1,...,k\}, \ S_l \leftarrow S_l \cap V $\\
	}
	Return $F$\\
	\caption{Approximation Algorithm for Directed Multi-Multiway Cut}
\end{algorithm}

\textbf{Lemma 3:} Algorithm 1 returns a Directed Multi-Multiway Cut.

\textbf{Proof:}  
For each ball like $B_x(s_{ij},r)$, where $1 \leq i \leq k, 1\leq j \leq |S_i|$, there is no vertex with the same group with $s_{ij}$ in $B_x(s_{ij},r)$ because the radius of each ball is smaller than $\frac{1}{2}$. The only case may lead to problems is that there are two vertices $ u $ and $ v $ in one ball, which are the members of another group and there is a path between them. In this case, only the central vertices and their incident edges will be deleted from the graph. Therefore, the path between the two vertices $ u $ and $ v $ will not be deleted from the graph. In the next iterations, at least one of the edges of the path between $ u $ and $ v $ will be put in the answer set.

\begin{flushright}
	$ \blacksquare $
\end{flushright}

\textbf{Theorem 1:} Algorithm 1 is a $(\alpha (1+\beta )k)$-approximation algorithm for Directed Multi-Multiway Cut.

\textbf{Proof:} According to Lemma 2, we have $c_i(r) \leq \alpha v_i(r) $, for every 
$ 1 \leq i \leq k$. Thus, $ \sum_{i=1}^{k} c_i(r) \leq \alpha \sum_{i=1}^{k} v_i(r) $. Besides, according to the definition of $v_i(r) $ and algorithm, we have
$ \sum_{i=1}^{k} v_i(r) \leq (kV^*+k\beta V^*) $ .Thus:

\begin{align*}
	F \leq \sum_{e \in F} w(e) = \sum_{i=1}^{k} c_i(r) \leq \alpha \sum_{i=1}^{k} v_i(r) \leq \alpha (1+\beta)kV^* \leq \alpha (1+\beta)kOPT.
\end{align*}

\begin{flushright}
	$ \blacksquare $
\end{flushright}

\textbf{The proof of Lemma 2:} We use the contradiction method. Assume that for every value of $r < \frac{1}{2}$ and every $\alpha$ $(\alpha > 0)$ we have 
$c_i(r)>\alpha v_i(r)$. Thus we have:

\begin{align*} 
	c_i(r) &> \alpha v_i(r) \\
	\int_{0}^{\frac{1}{2}} \frac{c_i(r)}{v_i(r)} dr &> \alpha \int_{0}^{\frac{1}{2}} dr\\
\end{align*}

According to Lemma 1, 
the function $v_i(r)$ is not differentiable at only a finite number of point. We call these points 
$r_0=0 \leq r_1 \leq ... \leq r_l \leq r_{l+1}=\frac{1}{2}$. Thus we have:

\begin{align*}
	\int_{0}^{\frac{1}{2}} \frac{1}{v_i(r)} (\frac{dv_i(r)}{dr})dr &= 
	\sum_{j=0}^{1}\int_{r_j}^{r_{j+1}} \frac{1}{v_i(r)}(\frac{dv_i(r)}{dr})dr \\
	& \leq ln v_i(\frac{1^-}{2})-ln v_i(0). \\
\end{align*}
Since $v_i(r)$ is an increasing function, we have 
\begin{align*}
	\leq lnv_i(\frac{1}{2})-lnv_i(0),
\end{align*}

as well $v_i(0)=\beta V^*$ as $v_i(\frac{1}{2})\leq \beta V^*+V^*$. Thus, we have:

\begin{align*}
	ln(\frac{\beta V^*+V^*}{\beta V^*}) &\geq ln(\frac{v_i(\frac{1}{2})}{v_i(0)})>\frac{\alpha}{2} \\
	ln(\frac{\beta+1}{\beta})&>\frac{\alpha}{2}. \tag{*}
\end{align*}

In order to reach a contradiction, we have to choose values for $\alpha$  and $\beta$  such that the inequality (*) will not be true. On the other hand, the approximation factor of the algorithm is dependent on these two parameters directly. So we have to choose the appropriate value for  $\alpha$  and $\beta$. Indeed, to find the best value for  $\alpha$  and $\beta$, we should solve the following nonlinear program:

\begin{align*}
	minimize \ \  &\alpha(1+\beta) \\
	subject\ to \ \  &ln(\frac{\beta+1}{\beta}) \leq \frac{\alpha}{2} \\
	& \ \  \alpha , \beta > 0.
\end{align*}

We have solved this nonlinear program using Matlab software and found the optimal value of $\alpha$  and $\beta$. These values are as follows $\alpha = 0.1$ and $ \beta = 20.504$. If we put these values in the inequality (*), the contradiction is reached and Lemma 2 is proved. Using these values for $\alpha$ and $\beta$, the algorithm is an approximation algorithm with factor $(2.1504)k$ for the Directed Multi-Multiway cut problem. 

\begin{flushright}
	$ \blacksquare $
\end{flushright}

\section{Approximation Algorithm for Directed Multicut}  \label{section.dm}
Similar to the LP presented in the previous section can be provided an LP for the Directed Multicut problem. 
We define a decision variable $x(e)$ for each edge $e$ which is as follows. If $e$ belongs to the directed multicut, $x(e)=1$, Otherwise $x(e)=0$. The purpose is to find a directed multicut with the minimum weight which cuts each directed path from $s_i$  to $t_i$  for $1 \leq i \leq k$. We represent the set of all directed paths from $s_i$  to $t_i$  for $ 1 \leq i \leq k$ with $P$. An linear programming for the problem is as follows:

\begin{equation*}
	\begin{array}{ll@{}ll}
		minimize  & \displaystyle\sum\limits_{e \in E} & w(e)x(e) &\\
		subject$ $to & \displaystyle\sum\limits_{e \in p} &x(e) \geq 1,  & \forall p \in P \\
		&                                                &x(e) \geq 0, & \forall e \in E.
	\end{array}
\end{equation*}

In this LP, there is a constraint for each path. On the other hand, we may have an exponential number of paths with respect to the input size and as a result, an exponential number of constraints. Nevertheless, we can solve this LP in polynomial time, using the ellipsoid algorithm. For this LP, the separation oracle operates as follows: we get a solution $x$ and assume that the length of each edge $e$ is equal to $x(e)$. Then, we find the shortest directed path from $s_i$  to $t_i$,  for $1 \leq i \leq k$. If the shortest path from $s_i$  to $t_i$, is more than 1, then the constraint 
$\sum_{e \in p} x(e) \geq 1$  is true for all paths from   $s_i$ to $t_i$. So this LP can be solved in polynomial time.

We provide a direct version of definitions like those used in region growth technique in \cite{1}. We define a distance on edges, assume that $x$ is an optimal solution for LP, let $x(e)$ be the length of edge $e$. We show the shortest path from $u$ to $v$, which is based on $x(e)$, with $dist(u,v)$. If there is not any directed path from $u$ to $v$, the value of $dist(u,v)$ is the shortest path between $u$ and $v$ in the graph, without noticing the direction of edges. Now we define:

\begin{align*}
	B_x(s_i,r)=\{v \in V : dist(s_i,v) \leq r\}.
\end{align*}

$B_x(s_i,r)$ is an area like a ball with center $s_i$, where $1 \leq i \leq k$, and radius $r \in \mathbb{R}$. 
Assume that the product of $w(e)x(e)$ is equal to the volume of edge $e$. Thus, the solution of the LP is the minimum volume of edges such that $dist(s_i,t_i) \geq 1$, for $1 \leq i \leq k$. Assume that $x$ is an optimal solution for the LP. Let $V^*=\sum_{e \in E} w(e)x(e)$  be the volume of all edges, indeed $V^*$ is the optimal value of LP. We know that $V^* \leq OPT$  such that $OPT$ is the optimal value for the IP. $v_x(s_i,r)$  is defined as follows:

\begin{align*}
	v_x(s_i,r) = \beta V^* + 
	& \sum_{
		\substack{
			e=(u,v)\in E \\ u,v \in B_x(s_{i},r)
		}
	} w(e)x(e)+ \sum_{
		\substack{
			e=(u,v) \in E \\ u \in B_x(s_{i},r) \\ v \notin B_x(s_{i},r)
		}
	} w(e)(r-dist(s_{i},u)).
\end{align*}

Let $ \delta(s)$ be the set of all edges which only one of their endpoints is in the set $s$. For a given radius $r$, we define:

\begin{align*}
	wt(\delta(B_x(s_{i},r)))=\sum_{e \in \delta(B_x(s_{i},r))} w(e).
\end{align*}

According to these definitions, we can express Lemma 4, which is used in the proof of Lemma 5.

\textbf{Lemma 4:} The function  $v_x(s_i,r)$  is differentiable in $(0,\infty)$  except some finite numbers of points. The derivative of this function is $wt(\delta(B_x(s_{i},r)))$.

Lemma 5 demonstrates that in directed graphs, we can always find a radius $r<\frac{1}{2}$, such that the cost $v_x(s_i,r)$ is an upper bound for $wt(\delta(B_x(s_{i},r)))$.

\textbf{Lemma 5:} Assume that $x$ is a feasible solution for  LP. For every $s_i$  there is a $r < \frac{1}{2}$ and at least an $\alpha$ $(\alpha>0)$ such that the following inequality is true:

\begin{align*}
	wt(\delta(B_x(s_{i},r))) \leq \alpha v_x(s_i,r).
\end{align*}

The proof of Lemma 4 and Lemma 5 is similar to the proof of Lemma 1 and Lemma 2, respectively.
In the rest of the paper, for simplicity we assume that $wt(r)=wt(\delta(B_x(s_{i},r)))$ and $v(r)=v_x(s_i,r)$.

Our polynomial time approximation algorithm for directed multicut is described in Algorithm 2.
The algorithm first solves the LP and finds the optimal solution $x$. In every iteration, the algorithm finds a pair which there is a path between and finds an area with a radius that satisfies the condition in Lemma 5. Then, the algorithm puts the edges, which have been cut by the area, in set $F$. If the area includes another pair $(s_j,t_j)$, and there is a path from $s_j$ to $t_j$, then the algorithm removes only the central vertex of the area. Therefore, we can cut the path between $s_j$ and $t_j$ in the next iterations.

\begin{algorithm} [H]
	\SetAlgoLined
	\KwResult{A Directed Multicut }
	$F \leftarrow \emptyset $ \\
	Solve the LP and get an optimal solution $x$ \\
	\While{there is a path between $s_i$ to $t_i$, where $1 \leq i \leq k$}{
		Grow a region $S=B_x(s_i,r)$ until $wt(r) \leq \alpha v(r)$\\
		Add $\delta(S)$ to F\\
		\eIf{ $S$ contains a pair $(s_j,t_j)$, where $1 \leq j \leq k$ and $i \neq j$ AND there is a path form $s_j$ to $t_j$}{
			Remove $s_i$ and incident edges with $s_i$ form the graph\\
		}{
			Remove $S$ and $\delta(S)$ form the graph\\
		}
	}
	Return $F$\\
	\caption{Approximation Algorithm for Directed Multicut}
\end{algorithm}

\textbf{Lemma 6:} Algorithm 2 returns a directed multicut.

\textbf{Proof:} Consider the ball $B_x(s_i,r)$. The center of this ball is $s_i$. Vertex $t_i$ cannot be in this ball because the radius of the ball is less than $\frac{1}{2}$ $(r<\frac{1}{2})$. Beside this, according to the LP constraint, we know that the distance between every pair $(s_i,t_i)$, where $1 \leq i \leq k$, should be more than 1. The only case may lead to problems is that there is a pair $(s_j,t_j)$, where $1 \leq j \leq k$ and $i \neq j$, in $B_x(s_i,r)$ such that there is a path from  $s_j$  to $t_j$. In this case, the algorithm only removes the central vertex $s_i$ from the graph, thus the pair $(s_j,t_j)$  and also the path between these vertices are still in the graph. In the rest of iterations, the algorithm will make them disconnected.
\begin{flushright}
	$ \blacksquare $
\end{flushright}

\textbf{Theorem 2:} Algorithm 2 is an $O(k)$-approximation algorithm for the Directed Multicut problem.

\textbf{Proof:} We demonstrate the set of vertices in the ball $B_x(s_i,r)$ with $B_i$. We assume that $B_i=\emptyset$  when no ball is selected for vertex $s_i$. We also demonstrate the set of cut edges for $B_i$  with $F_i$. It means $F_i$ is equal to $\delta (B_i) $. Thus, we have $F=\bigcup_{i=1}^{k} F_i$. Assume that $V_i$ is equal to the volume of all edges which are in the ball $B_i$ and also the volume of edges which have one head in $B_i$. According to this definition, we have $V_i \geq v_x(s_i,r)-\beta V^*$ because $V_i$ includes the volume of all edges in $F_i$. But $v_x(s_i,r)$ is contained only some part of these edges and an addition value $\beta V^*$.
According to Lemma 5 and the value chosen for $r$ in the algorithm, we have $wt(F_i) \leq \alpha v_x(s_i,r) \leq \alpha (V_i + \beta V^*)$. We know that the algorithm may not remove the edges which are incident with vertices in $B_i$ in this iteration. Thus, an edge may belong to more than one area, on the other hand, there are at most $k$ areas. Therefore, $\sum_{i=1}^{k} V_i \leq k V^*$. So we have the following inequalities:

\begin{align*}
	\sum_{e \in F} w(e)=\sum_{i=1}^{k} wt(F_i) \leq \alpha \sum_{i=1}^{k} (V_i+\beta V^*) \leq \alpha(1+\beta)kV^* \leq \alpha (1+\beta)kOPT.
\end{align*}

\begin{flushright}
	$ \blacksquare $
\end{flushright}

Similar to section \ref{section.dmmc}, the optimal value for $\alpha$ is $0.1$ and $\beta$ is $20.504$. So Algorithm 2 is an $(2.1504)k$-approximation algorithm for the Directed Multicut problem. 

\section{Conclusions} \label{section.con}
In this paper, we design approximation algorithms for the directed multi-multiway cut and directed multicut problems using the region growing technique \cite{1,2}. By this paradigm, we give for each problem an $O(k)$-approximation algorithm. The works previously done on these problems need to solve k linear programs, whereas our algorithms require only one linear programming. Both algorithms use the same linear programming relaxation. A question of interest is to find the integrality gap of the linear programming relaxation for these problems.

\end{document}